**Time-resolved Imaging of Stochastic Cascade Reactions over a Submillisecond to Second Time Range at the Angstrom Level**

Toshiki Shimizu,[1,†] Dominik Lungerich,[1,2,3,*] Koji Harano,[1,*,#] and Eiichi Nakamura[1,*]

[1] Department of Chemistry, The University of Tokyo, 7-3-1 Hongo, Bunkyo-ku, Tokyo 113-0033, Japan

[2] Center for Nanomedicine (CNM), Institute for Basic Science (IBS), IBS Hall, 50 Yonsei-ro, Seodaemun-gu, Seoul, 03722, South Korea.

[3] Graduate Program of Nano Biomedical Engineering (NanoBME), Advanced Science Institute, Yonsei University, Seoul, 03722, South Korea.

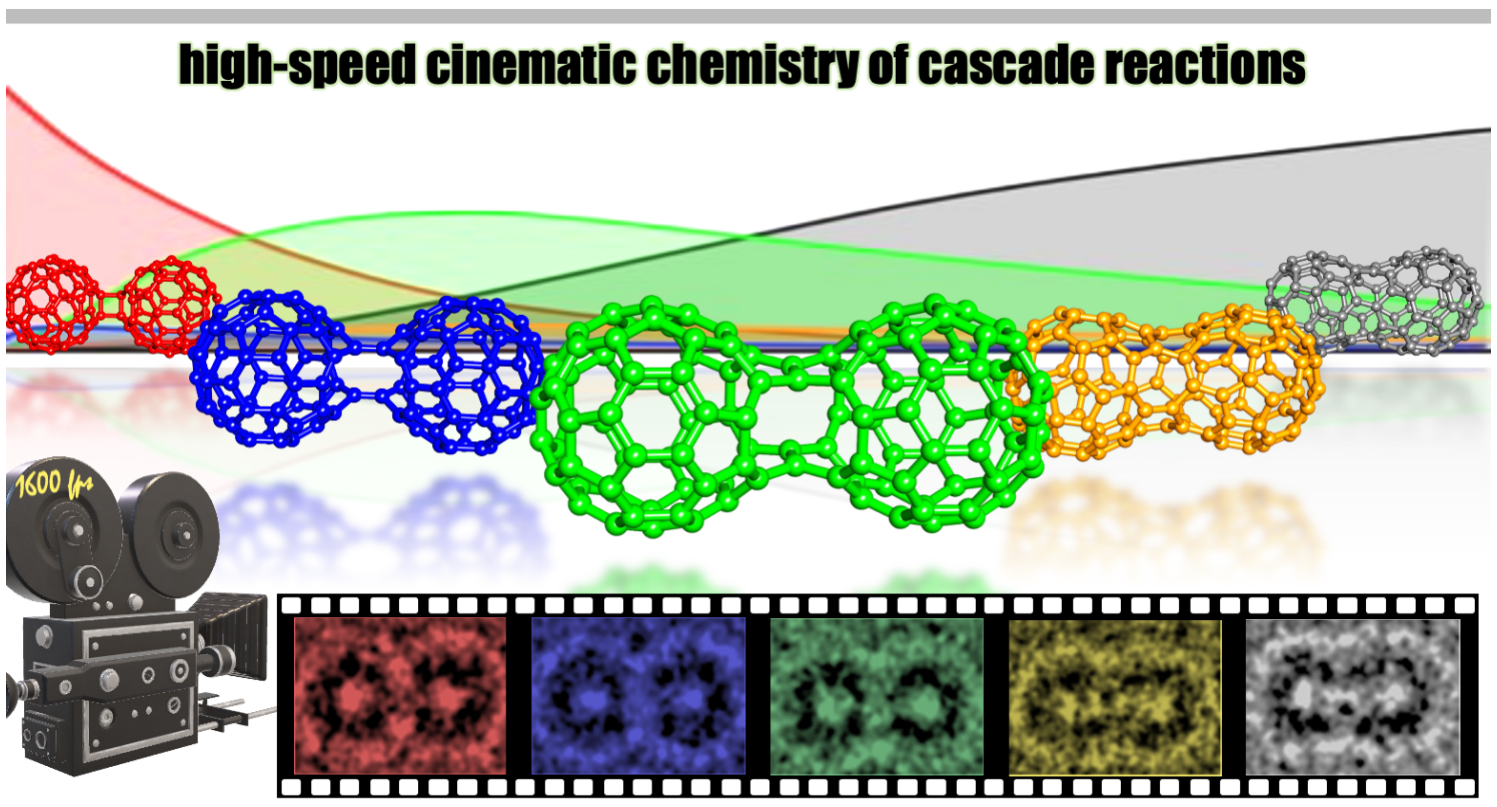

**ABSTRACT:** Most chemical reactions are cascade reactions in which a series of transient intermediates appear and disappear stochastically over an extended period. The mechanisms of such reactions are challenging to study, even in ultrafast pump–probe experiments. The dimerization of a van der Waals dimer of [60]fullerene producing a short carbon nanotube is a typical cascade reaction and is probably the most frequently studied in carbon materials chemistry. As many as 23 intermediates were predicted by theory, but only the first stable one has been verified experimentally. With the aid of fast electron microscopy, we obtained cinematographic recordings of individual molecules at a maximum frame rate of 1600 frames per second. Using Chambolle total variation algorithm processing and automated cross-correlation image matching analysis, we report on the identification of several metastable intermediates by their shape and size. Although the reaction events occurred stochastically, varying the lifetime of each intermediate accordingly, the average lifetime for each intermediate structure could be obtained from statistical analysis of many cinematographic images for the cascade reaction. Among the shortest-living intermediates, we detected one that lasted for less than 3 ms in three independent cascade reactions. We anticipate that the rapid technological development of microscopy and image processing will soon initiate an era of cinematographic studies of chemical reactions.

## 1. INTRODUCTION

The dimerization of a [60]fullerene ($C_{60}$) van der Waals (vdW) dimer forming a short carbon nanotube (CNT) is probably the most frequently studied transformation in the field of carbon chemistry and has been observed under high pressure,[1] high temperature,[2,3] metal reduction,[4] light and electron irradiation,[5,6,7,8,9] as well as mechanochemical conditions.[10,11,12] Osawa and Tomanek proposed by theory the shortest possible path of this transformation from the vdW dimer **OT-0** (Figure 1a), which is initiated by a cycloaddition/retrocycloaddition (Figure 1b), followed by a series of Stone–Wales rearrangements[13,14] (Figure 1c) to a $D_{5d}$ symmetry (5,5) CNT **OT-24**.[15,16] Although the structure of the first long-lived intermediate, the [2 + 2] cycloadduct (**OT-1**), was identified using X-ray diffraction[10] and by high-resolution transmission electron microscopy (TEM),[17] no other structures have been experimentally verified because of either their short lifetimes or the structural inhomogeneity and complexity of the intermediates. Typical for such fast cascade reactions are the many temporally overlapping transient intermediates, which appear and disappear stochastically during the transformation (Figure 1d). Hence, following the time course in bulk experiments becomes difficult because of overlapping signals of each intermediate that may also be present in low concentration (Figure 1e). These challenges are also true for ultrafast pump–probe experiments, which require the observed events to be repeatable and easily reproducible.[18]

Generally speaking, high energy reactive species often play a more critical role in chemical processes than low-energy species. However, they are more short-lived and hence more difficult to detect experimentally. Nonetheless, the analysis becomes possible if the method continuously monitors individual molecules and their reactions.[19,20,21,22,23] We recently reported the cinematographic study of converting a $C_{60}H_{30}$ molecule into fullerene $C_{60}$ at 2–25 frames per second (fps) over tens of seconds,[24] using single-molecule atomic-resolution time-resolved electron microscopy (SMART-EM).[25,26,27,28] In this work, we studied the cascade dimerization of single vdW dimers of $C_{60}$ using a direct electron detector (DED) camera operating at 1600 fps, which expanded the analytical time range of the study by a factor of $10^5$. Through the use of a Chambolle total variation (CTV) denoising algorithm, a method widely employed for the denoising of optical digital images,[29,30,31] and automated cross-correlation (XC) image matching analysis of density functional theory (DFT)-based TEM-simulations,[24,32] we report here the experimental identification of **OT-0**, **1**, **2**, **4**, **11**, and **14** intermediates, which were distinguished by their shape and size, and we could resolve the lifetime of individual intermediates under our standard TEM conditions. The reaction events were purely stochastic, and the lifetime of each metastable intermediate also changed stochastically. Among these intermediates, we detected short-lived species **OT-2** (<3 ms), detected only for three cases in several tens of cascade reactions from several minutes of cinematographic datasets (see Supporting Information Videos S1–S4).

Further, the lifetimes varied widely, as exemplified for vdW dimer **OT-0**, which lived between 1 and 150 s at a standard electron dose rate (EDR) of $2.2 \times 10^6$ $e^-$ $nm^{-2}$ $s^{-1}$, as examined from several tens of cases (Figure 2a).[33] Similarly, **OT-1** and **OT-14** lived for 1 to 20 s (Figure 2b), and **OT-4** and **OT-11** for 20 to 300 ms each (Figure 2c). In our best effort at sorting out nearly a million 0.625-ms frames, we could determine the timing of a single reaction event with a precision of 1.25 ms at the shortest. From a distance dataset, we obtained the average distance, *d*, with a standard error of 0.02–0.04 nm by second-scale imaging with a confidence level of 99% (Figure 2b), and the localization precision becomes even better (0.01–0.02 nm) at millisecond-scale imaging providing many video frames (Figure 2c). Our data not only provide experimental support for the **OT** diagram (Figure 1a) but illustrates the power of cinematographic techniques such as SMART-EM to study molecular events with submillisecond and subangstrom precision.[34]

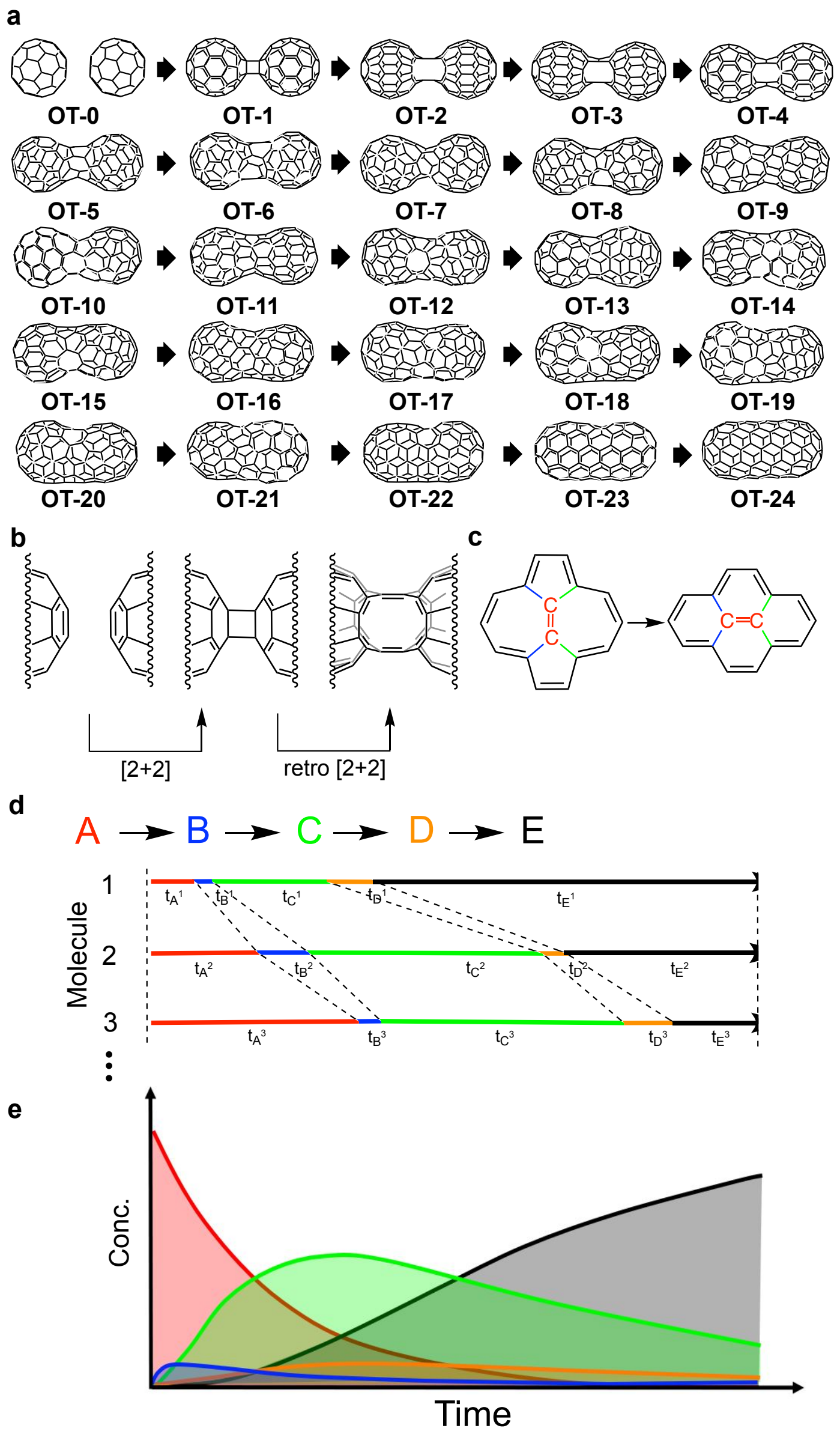

**Figure 1.** Cascade reaction and conversion of $C_{60}$ vdW dimer to a short carbon nanotube. (a) A proposed pathway for converting a $C_{60}$ vdW dimer (**OT-0**) to the short $C_{120}$ nanotube (**OT-24**), involving the smallest number of Stone–Wales rearrangements. (b) [2 + 2] and retro [2 + 2] reactions of the $C_{60}$ dimer. (c) Stone–Wales rearrangement to convert two pairs of 7- and 5-membered rings to four 6-membered rings. (d) Exemplified multistep conversion of intermediates A to E and the lifetimes of individual molecules. (e) Concentration of the intermediates as the sum of all molecules

in the system. Short-lived intermediates (blue and yellow) are challenging to detect due to their low concentration throughout the cascade.

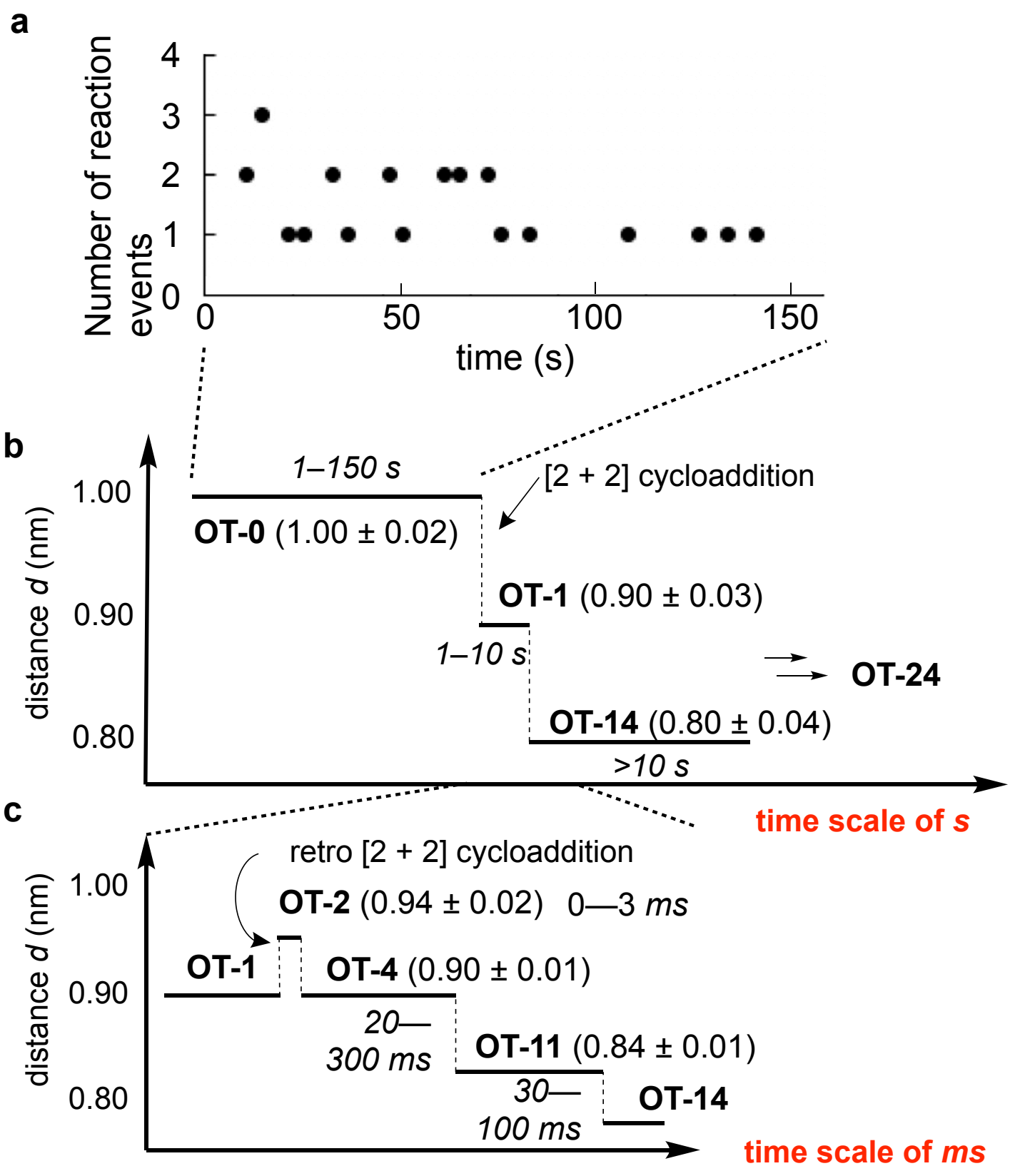

**Figure 2**. Time evolution of the conversion of $C_{60}$ vdW dimers to short CNTs. (a) Time-dependent change of the dimerization frequency via the [2 + 2] cycloaddition at 80 kV results in the shortening of the center-of-mass distance between two $C_{60}$ entities from 1.0 nm to 0.9 nm. Taken from ref 33. Lifetimes under the standard EDR of 2.2 × $10^6$ $e^-$ $nm^{-2}$ $s^{-1}$ are shown. (b) Time course of the change of the distance *d* on the time scale of seconds. (c) Time course of the conversion of **OT-1** to **OT-14** on the time course of milliseconds. Errors of the distance *d* are the standard errors of the mean at the 99% confidence level.

## 2. METHODS

We used a Cs aberration-corrected JEOL JEM-ARM200F transmission electron microscope equipped with a DED camera (Gatan K2-IS). SMART-EM imaging was conducted with an EDR between 2.2 and $9.5 \times 10^6$ $e^- nm^{-2} s^{-1}$, and an acceleration voltage of 80 kV for high-contrast imaging at the high frame rate of 1600 fps (exposure time 0.625 ms/frame). The reaction rate was previously found to be only dependent on the total electron dose but not on the EDR.[19] $C_{60}$@CNT samples were prepared as previously published.[19] We first surveyed many hundreds of $C_{60}$ vdW dimers (i.e., **OT-0**) encapsulated in CNTs at low EDR (<$10^4$ $e^- nm^{-2} s^{-1}$) to find suitable molecules in CNT bundles for cinematographic imaging of reactions. Upon finding suitable candidates, we blanked the beam for 1 min to minimize the thermal drift of the specimen. Then, we set the EDR to the target value and started the recording. The defocus value was set manually to ca –20 nm (underfocus), which is larger than typical SMART-EM experiments, to enhance the molecular contrast so that we can detect the molecules in the high frame rate experiments. All images were processed using a bandpass filter (filtering structures smaller than 3 pixels and larger than 40 pixels, tolerance of direction: 5%) and adjusted with brightness and contrast.

Raw images recorded at 1600 fps are full of shot noise, and the $C_{60}$ molecules are hardly discernable. Therefore, for cinematographic tracking of the dimerization reactions, we employed CTV denoising,[29] which removes noise while preserving the original signal that contains the chemical and morphological information of the specimen. We set a weight parameter of 0.2–0.5 as a compromise between the signal-to-noise ratio and localization precision, which was ±0.03 nm under the worst imaging conditions. To overcome grouped shot noise signals using CTV, we superimposed as many as necessary, but as few as possible frames, for the best possible temporal resolution in each scene (compare Figure 3a). The processed videos are shown as Supporting Information Videos S1–S4.

We determined the center of mass of each $C_{60}$ entity along the tube axis (red line, Figure 3b) and determined the distance $d$ with a statistical error of 0.01 nm. From statistical analysis,[35] we obtained a mean value of distance $d$, with the standard deviation $\sigma$, and the standard error ($\sigma_x$). An example of a series of TEM images (3.125 ms/frame) for the evolution of the fusion is shown in Figure 3c. In some cases, we could not measure *d* because of low image quality caused by either the reaction, specimen vibration, or noise. The time-course–distance analysis did not include frames without a distance due to blurring.

Structural assignment of the intermediates was carried out by comparing the experimental distances and shapes of TEM image simulations of molecular models obtained from DFT calculations at the ωB97X-D/def2-TZVP//6-31G(d) level of theory. Because most **OT** intermediates are $C_1$-symmetric, we derived a simulation library of rotated intermediates along the longitudinal axis between 0° and 350° in 10° steps (Figure 4a). Then, we determined the center-of-mass distance from the periphery of the fused dimers as indicated in Figure 4b and obtained the mean distance, *d*, and the standard deviation for the 36 rotational simulations.

The similarity between the experimental TEM images and 864 simulations (24 intermediates from 36 different directions) was quantified using XC factors[36] between the simulation and experimental images automatically obtained using in-house-developed XC software (Figure 4c). The XC factor was typically close to 0.7 for the best match and 0.4 for the worst. Details are shown in Supporting Information.

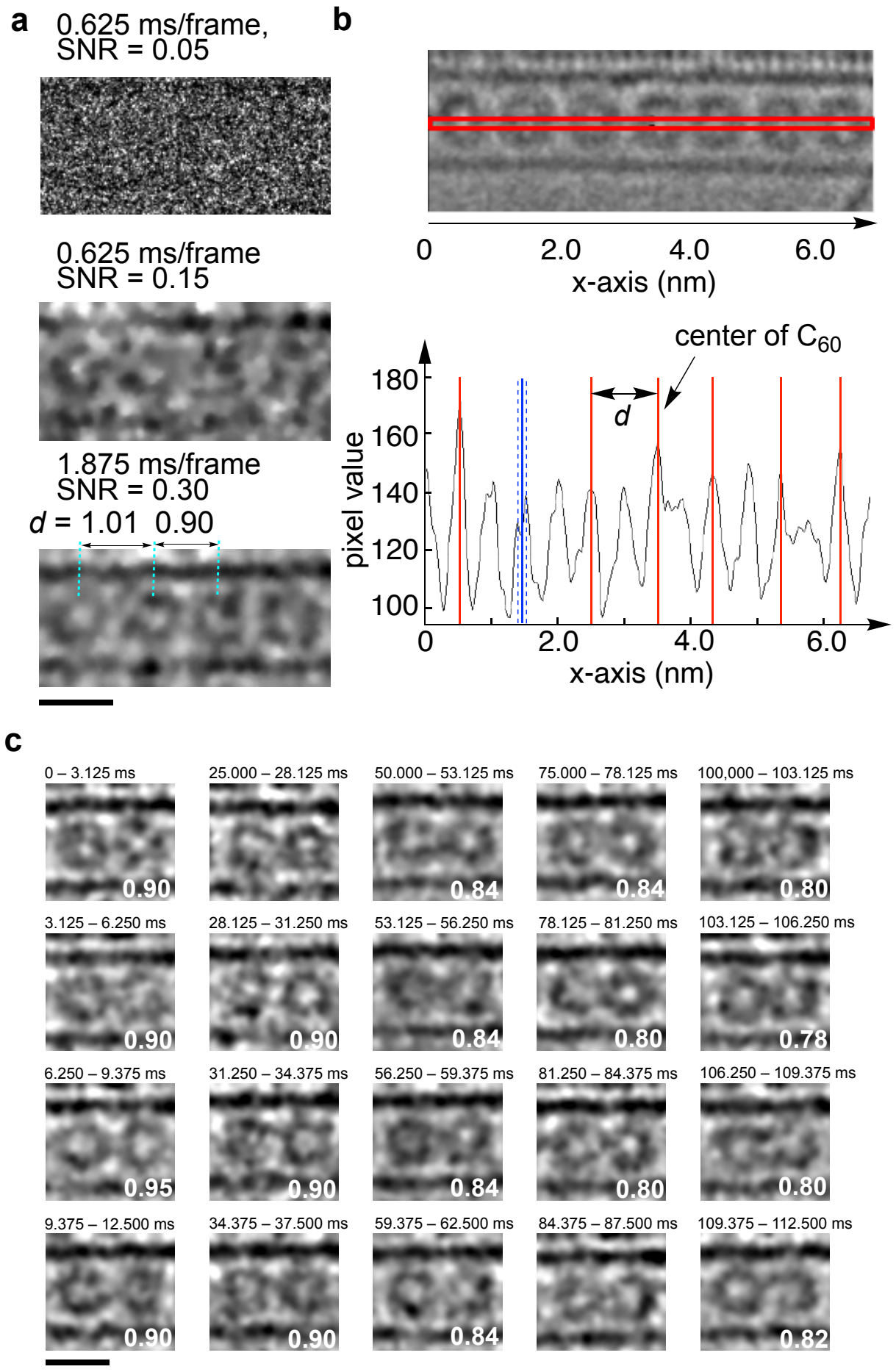

**Figure 3**. Image processing and distance analysis for ultrafast SMART-EM video frames for the dimerization reaction of $C_{60}$. (a) Exemplified 1600 fps raw image, after CTV denoising, and after superimposition of three frames. Scale bar: 1 nm. (b) Determination of the center-of-mass distance *d* between two $C_{60}$ molecules. The selected area for distance analysis is shown in the red box, which results in the intensity profile along the *x*-axis (below). (c) Cinematographic frames (3.125 ms/frame) of a $C_{60}$ dimer, undergoing the multistep conversion in a CNT. The time begins with the start of recording, and numbers on the right bottom are the intermolecular distances, *d*, in nm. See Supporting Information Video S1 for the corresponding video images.

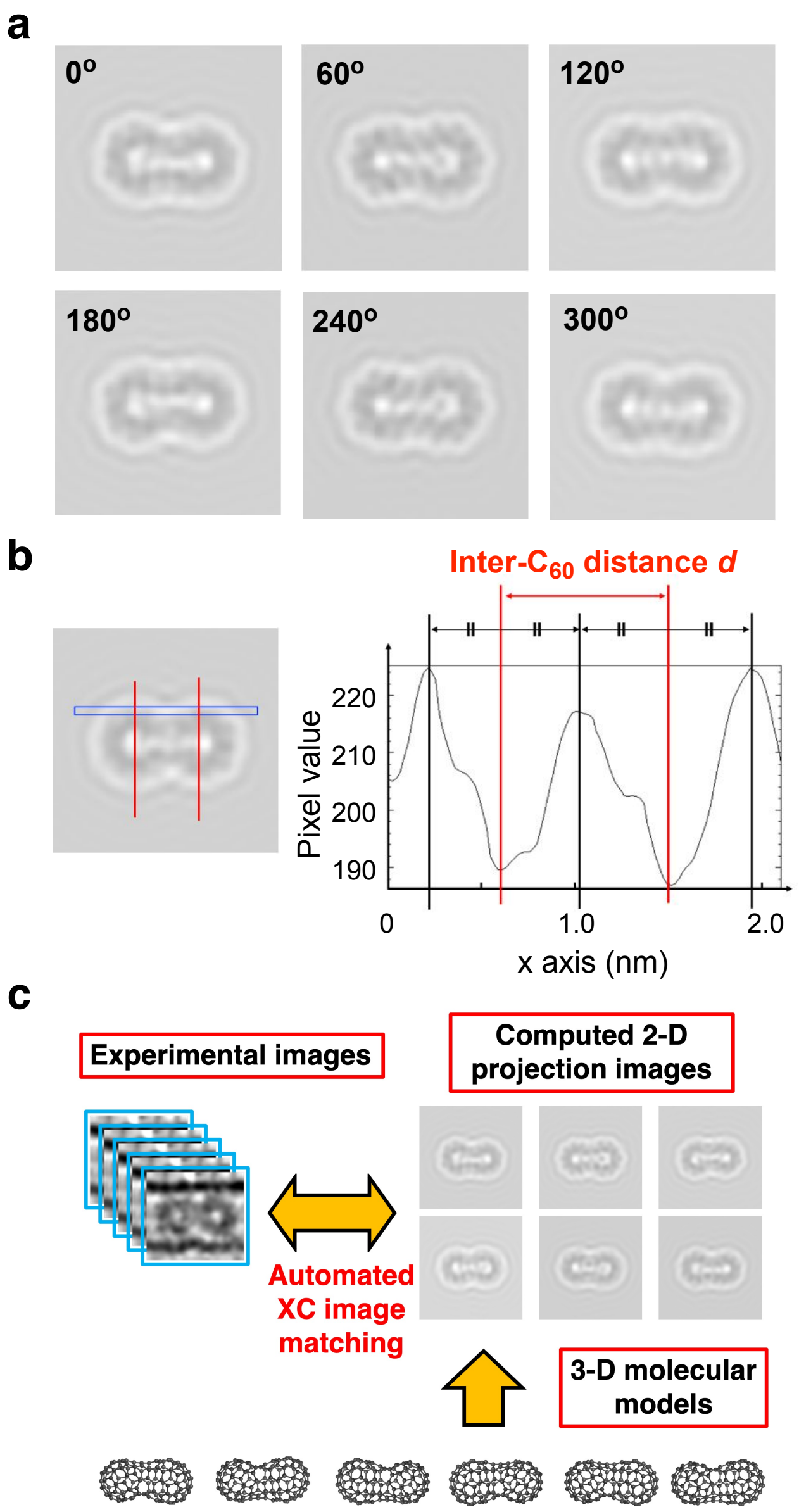

**Figure 4.** Structure assignment of intermediates by distance (*d*) measurement and cross-correlation (XC) image matching analysis with TEM simulation images. (a) A series of TEM simulation images of **OT-14** rotated along the longitudinal axis by 60° (see all 36 images rotated by 10° steps in Supporting Information). (b) Intensity profile of the selected area (blue box) in the TEM simulation image of **OT-4** to obtain *d* as a distance between local minima of the pixel value plot. (c) Schematic illustration of a quantitative assessment of similarities between experimental and simulated TEM images via XC.

## 3. RESULTS AND DISCUSSION

**3.1. Precise Time Determination of Reaction Occurrence.** We first describe how precisely we can determine the time of the occurrence of each reaction event for the conversion of **OT-0** to **OT-1**. For these stable intermediates, we superimposed many images to determine precisely the change of the distance *d* between two molecules (compare Supporting Information). Figure 5a shows four images taken from a 62.5 ms/frame video over 250.0 ms after an arbitrary time 0. The distances for the molecular pairs 1–2, 3–4, 4–5, and 5–6 are between 0.97 and 1.08 nm, indicating that the $C_{60}$ molecules are in vdW contact with each other or a little farther away. The average distance *d* measured for 30 different vdW contacts at 423 K was 1.00 ± 0.01 nm with a confidence level of 99%, comparing favorably with the X-ray crystallographic data for *d* of 1.002 ± 0.001 nm at 300 K.[37] This result guarantees that the measuring method described here is accurate enough for quantitative analysis of the distance. The 2–3 distance is 0.80 nm, indicating that the two molecules have already reacted before the start of the observation. This intermediate is often qualitatively described in the literature as a "peanut-shaped" dimer, which we assign to be **OT-14** (*vide infra*). An important observation is that the 4–5 distance suddenly decreased from 1.01 nm to 0.91 nm between the 62.5-ms and the 125.0-ms frame, indicating that [2 + 2] cycloaddition took place around this time. The distance averaged over the same cycloadduct in a series of video frames was 0.88 ± 0.02 nm, which agrees with the X-ray crystallographic data of 0.884 ± 0.001 nm.[10]

To determine when the cycloaddition took place, we examined the details of the 125.5–187.5-ms frame to obtain the noisy raw 160 fps images shown in Figure 5b. The Chambolle denoising afforded cleaner pictures in Figure 5c, where we can see that the 4–5 distance contracted between the 122.5–135.0-ms frames. Further temporal deconvolution to 4.375 ms/frame (Figure 5d), to 1.875 ms/frame (Figure 5e), and finally, to 1.25 ms/frame images in Figure 5f narrows down that the [2 + 2] cycloaddition took place between the two frames, 127.5–128.8 ms and 128.8–130.0 ms. As summarized in Figure 5 for the time course of the reaction measured at frame rates from 6.25 to 1.25 ms, we can reliably determine the time of a reaction event with a temporal precision of up to 1.25 ms using the present setup and the denoising algorithm. The average distance *d* for the same cycloadduct in the 1.25 ms/frame video was 0.91 ± 0.02 nm, which agrees with that obtained from the X-ray structure, indicating the high precision of distance measurement even at the 1.25-ms imaging.

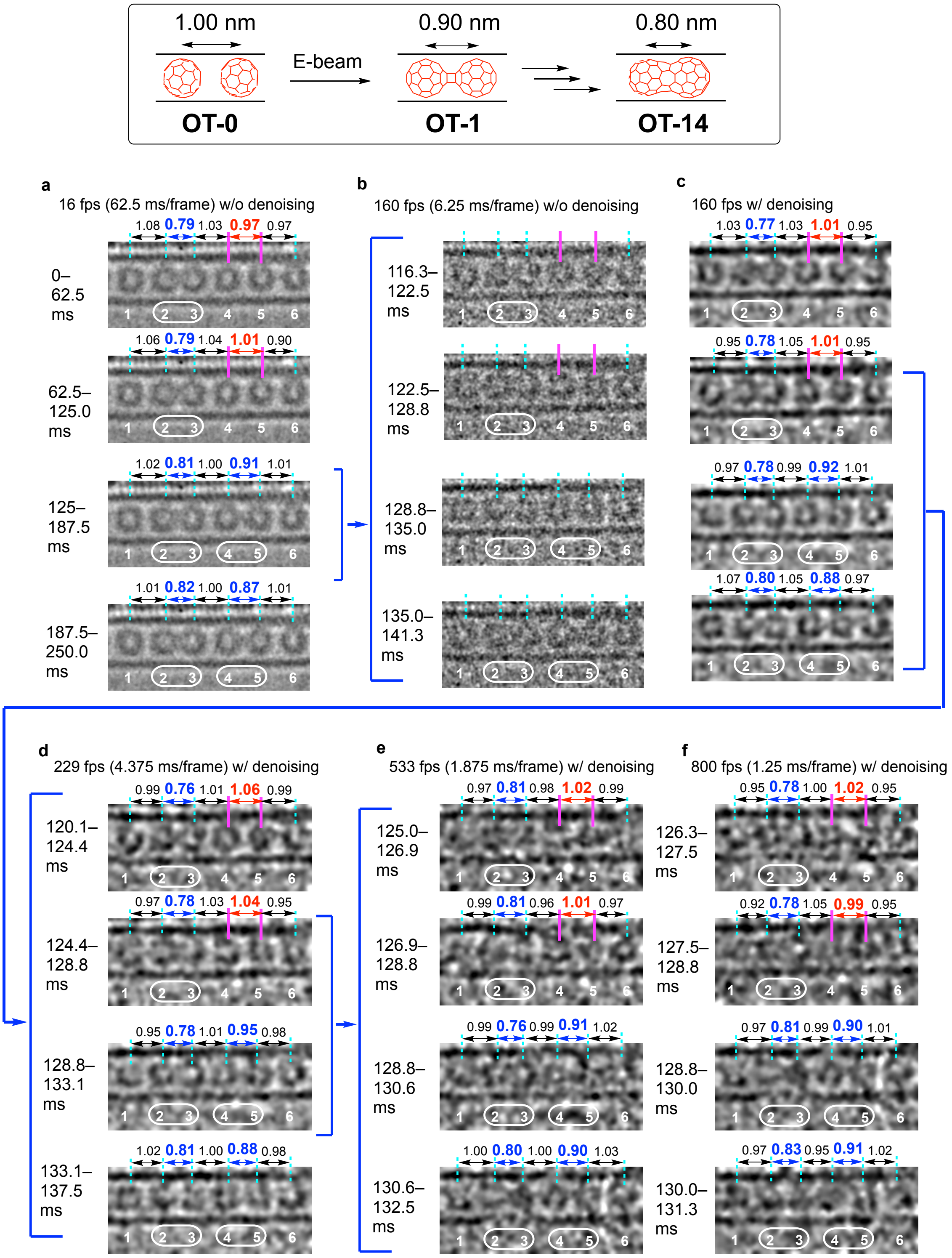

**Figure 5.** Frames showing a cycloaddition event ($C_{60}$ molecules **4** and **5**) in a CNT at 423 K (80 kV, EDR = $2.2 \times 10^6$ $e^-$ $nm^{-2}$ $s^{-1}$). $C_{60}$ molecules are numbered below, and dimers are circled together. Numbers above indicate the distance *d* in nm, measured for a stack of frames; the 0.021 nm pixel size ultimately limits the precision. Time 0 was set arbitrarily. (a–f) Frames of CTV denoised ((a) and (b) without denoising) and superimposed images with increasing temporal resolution. See also Supporting Information Video S2 for the corresponding video images.

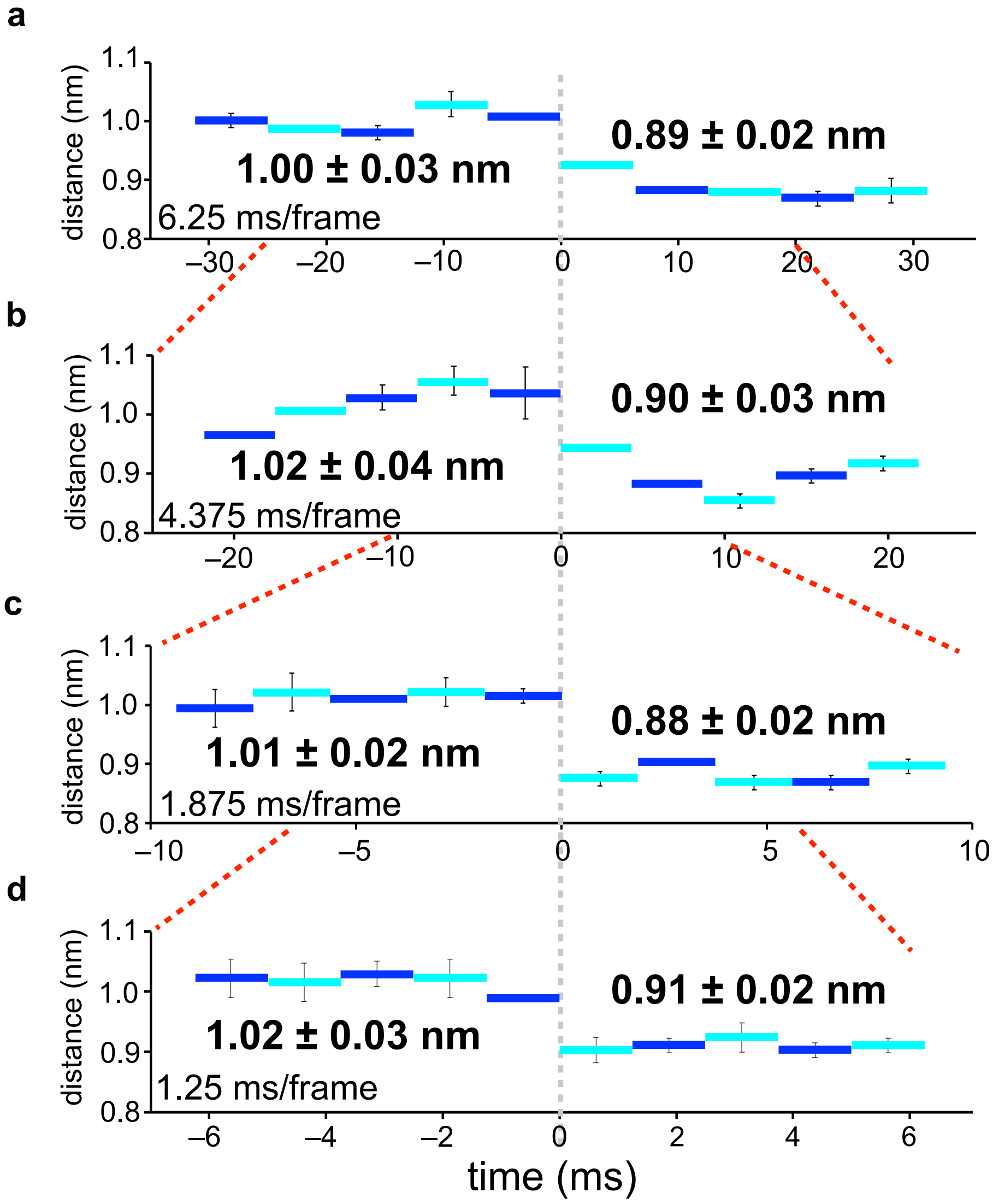

**Figure 6.** Determination of the time of the **OT-0** to **OT-1** conversion using a variety of frame rates. Time 0 is set to time 128.0 ms in Figure 5.

**3.2. Identification of Transient Intermediates of the $C_{60}$ Fusion Reaction.** Next, we utilized the above information to study when each intermediate appears and disappears during the conversion of **OT-0** to fused dimers. The Osawa–Tomanek pathway is remarkable in two aspects (Figure 1a). First, the retro [2 + 2] cycloaddition of **OT-1** produces a strained 12-membered ring in **OT-2** (Figure 1c) that results in the highest increase in energy ($\Delta E$, Figure 7a) and the longest $C_{60}$–$C_{60}$ distance ($d$ = 0.94 nm) (Figure 7b). Then, along the path, $d$ decreases constantly, accompanied by a subsequent energetic stabilization after **OT-6**. Note that the error bars for $d$ come from the distance variation in the two-dimensional projection upon rotation along the longitudinal axis of the (unsymmetrical) molecules (see Methods Section and Supporting Information). Second, the changes in energy ($\Delta E$) along the reaction coordinate in Figure 7a suggest the possibility of detecting thermodynamically trapped intermediates in local minima preceding high energy ones; these are **OT-1**, **4**, **11**, **14**, and **18**. On the other hand, **OT-2** requires a large amount of energy (284 kJ/mol) for transformation from **OT-1**, and such a high energy intermediate is expected to be hardly detectable in the cascade reaction. However, we noted that due to the increased assimilation to CNT **OT-24** and concomitant oligomerization/polymerization of the dimer,[19] the dimeric molecules after **OT-14** are difficult to define by $d$ and almost indistinguishable by their shape. Therefore, we limited our study to intermediates up to **OT-14**.

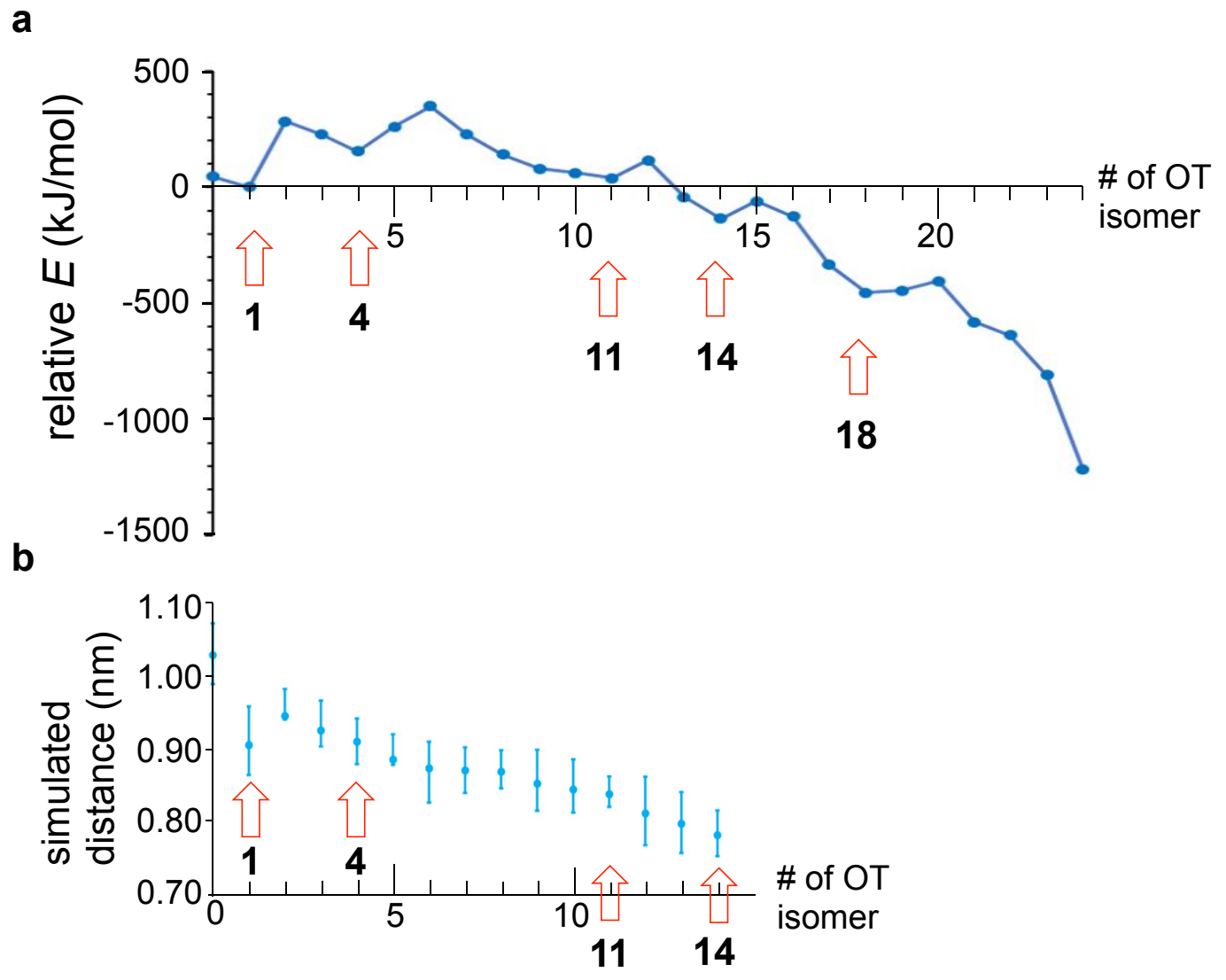

**Figure 7.** Energy diagram and simulated distance analysis of **OT** isomers. (a) Relative ground state energies of all **OT** isomers. The energy of **OT-1** is set to 0 eV. (b) Average of simulated distance *d* from **OT-0** to **OT-14**. Error bars are standard deviations derived from different angles.

Figure 8 shows the time-dependent conversion of **OT-0** to **OT-14** at 423 K as a function of the distance *d* measured every 3.125 ms. The dataset is shown in Supporting Information Video S1 and Figure S18. The *x*-axis in milliseconds starts from an arbitrary time 0 because **OT-0** and **OT-1** lived longer than 1 s (see Figure 2a,b). The structure of the [2 + 2] cycloadduct **OT-1** was previously determined unambiguously. The measured distance (0.90 nm) agrees with the reported distance (see above), and the XC value is high (0.641). Notably, after the formation of **OT-1**, the distance first increased to 0.95 nm in the 6.250–9.375-ms image before it decreased in approximately three steps via 0.90 nm, 0.84 nm, and 0.80 nm. With a measured $d = 0.95$ nm, **OT-2** is uniquely long and shows a very short lifetime of $<3$ ms; however, the high XC value of 0.671 supports the assignment. Note that we observed the **OT-2** intermediate only for three molecules out of several tens of molecules investigated by video imaging. The data fluctuations around 15, 88, 90–120 ms are likely errors due to CTV denoising artifacts (see above), pixel size, molecular motions, or vibration of the CNT.

**OT-4** is a species that lies in the bottom of a local energy minimum (Figure 7a) and is thus expected to have a relatively long lifetime, as found experimentally, appearing between ≈10 ms to 45 ms with $d = 0.90$ nm. Next, we assigned the species with $d = 0.84$ nm to be **OT-11**; however, the image may originate from an overlap of species ranging from **OT-8** to **OT-11**, which are almost indistinguishable by TEM. The species appearing at 78 ms is a long-lived species with a lifetime of up to 40 ms. Therefore, we assign it to **OT-14**, a local minimum with a calculated distance of $d =$ 0.80 nm, thus a perfect match with the experimental value. The exceptionally long lifetime must be why this species with $d = 0.80$ nm has been frequently reported in the literature as a "peanut-shaped" dimer,[2,17,38,39,40,41] along with the vdW dimer **OT-0** and the [2 + 2] cycloadduct **OT-1**; however, its exact structure was never determined.

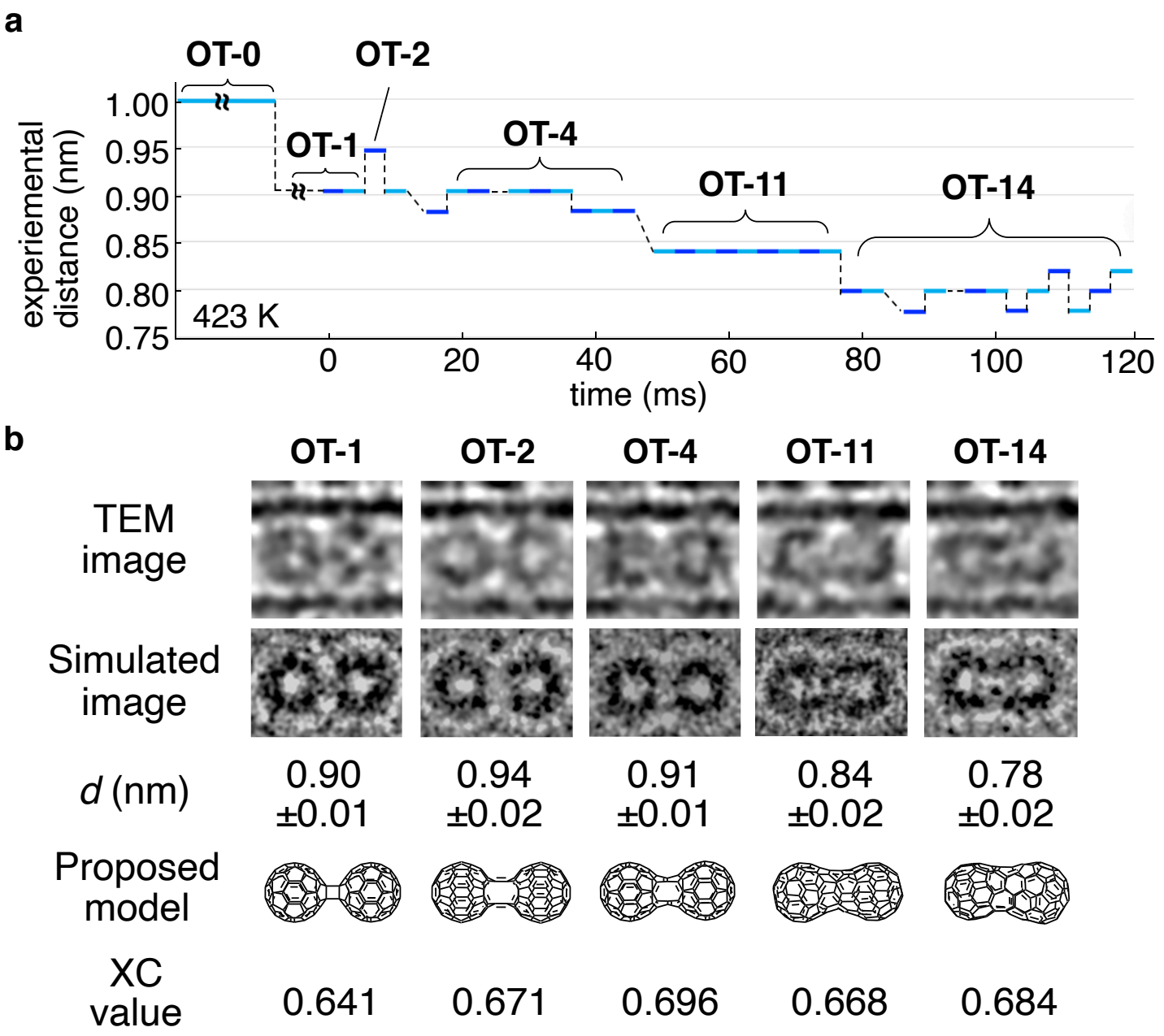

**Figure 8.** Cinematographic high-speed deconvolution of the multistep conversion of [2 + 2] cycloadduct **OT-1** to fused dimers at 423 K (80 kV, EDR = $2.2 \times 10^6$ $e^-$ $nm^{-2}$ $s^{-1}$). (a) Time evolution of distance *d* measured every 3.125 ms. To increase the visibility in the figure, each frame is shown alternately in blue and light blue. (b) Experimental and simulated images and *d* of observed intermediates, including the obtained XC values.

Finally, Figure 9 summarizes two more examples for molecules that reacted from **OT-0** to **OT-14** in a period of 120 to >2000 ms, which were recorded at 298 K (see Supporting Information Videos S3 and S4, as well as Figures S19 and S20 for frame images). Now, three key aspects become evident: First, in both additional examples in the cascade from **OT-0** to **OT-14**, we identified **OT-2**, which ensures that this short-lived intermediate in Figure 8 is not a misinterpretation of an artifact. The detection of **OT-2** suggests it to be a kinetically trapped intermediate with a relatively high energy barrier between **OT-2** and **OT-3**. Second, as a result of the variable lifetimes of each metastable intermediate, the reaction events occurred purely stochastically. In our best effort of sorting out nearly a million 0.625-ms frames, we could determine the timing of a single reaction event with a precision of 1.25 ms at the shortest, reaching the technological limitations of current state-of-the-art high-speed imaging. Third, at both temperatures, 298 K and 423 K, we detected the same sequence of intermediates, suggesting the proposed pathway by Osawa and Tomanek holds true at both temperature regimes. The average lifetimes for the intermediates obtained from a series of cinematographic experiments are summarized in Figure 2c.

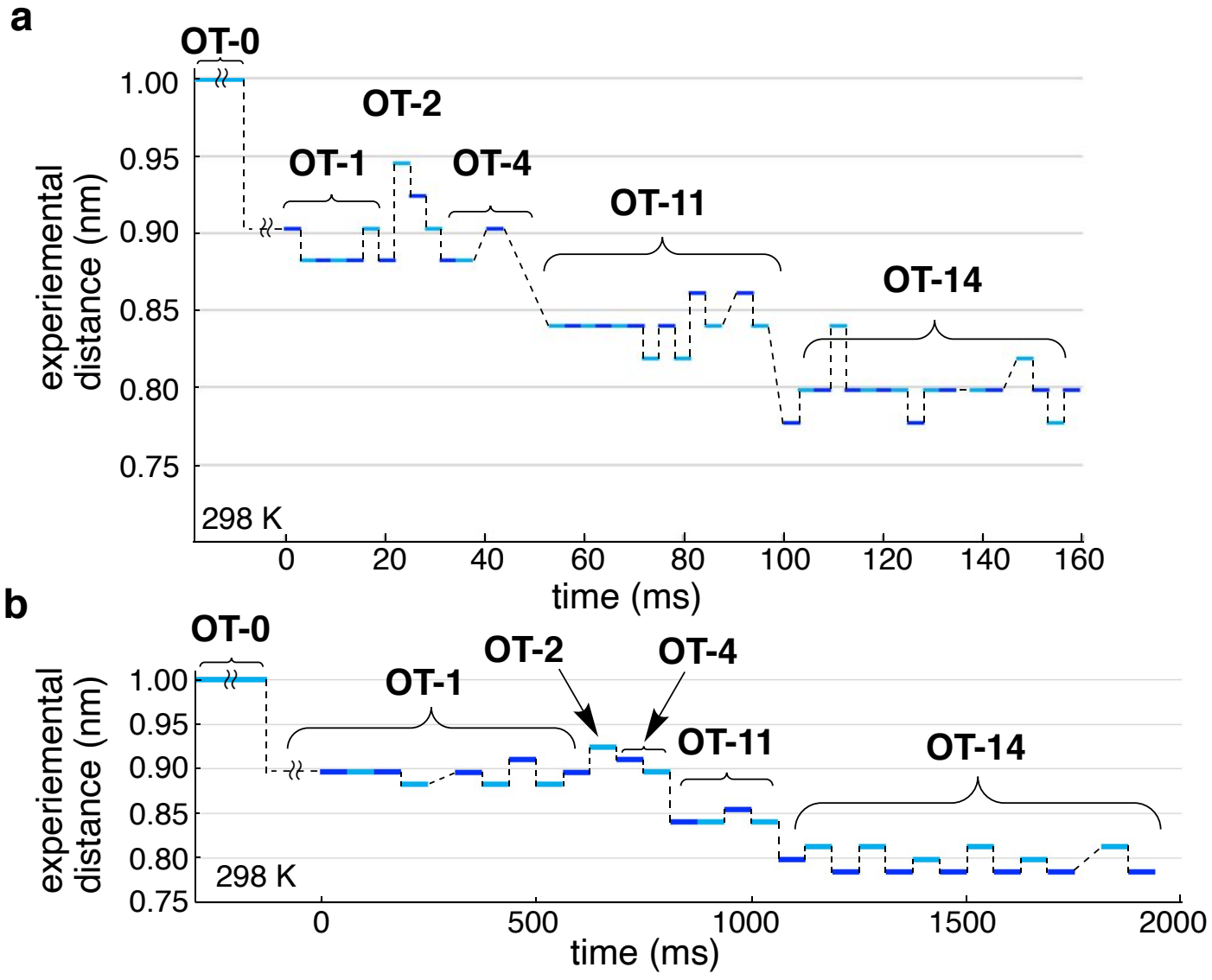

**Figure 9.** Deconvolution of the multistep conversion of [2 + 2] cycloadduct **OT-1** to fused dimers at 298 K (80 kV, EDR = $9.5 \times 10^6$ $e^-$ $nm^{-2}$ $s^{-1}$). (a) 3.125 ms frames. (b) 62.5 ms frames. See Supporting Information Videos S3 and S4 for the video images corresponding to a and b, respectively.

## CONCLUSION

In summary, we demonstrated by means of high-speed TEM imaging the power to deconvolute from cinematographic results the progress of the multistep cascade reaction between two $C_{60}$ molecules with unprecedented precision. So far, no attempts at their detection had been reported. We approached the problem by comparing the experimental and theoretical sizes of the molecules and succeeded in identifying four new intermediates, which were previously only proposed by theory. Particularly interesting is the detection of retrocycloadduct **OT-2**, which was identified from only a very short-lived time frame of around 3 ms and is an excellent example of an elusive high energy intermediate in a sequential reaction, which has never been detected experimentally before. The transformation of nanocarbon materials in the interior of a CNT is a model for the formation of noncrystalline regions in the interstices of crystalline graphitic regions in carbon fibers, providing experimental information on atomic-scale dynamics in the fiber formation.[42] Further, statistical analyses of several cascade reactions show that each intermediate appears and vanishes stochastically, meaning the lifetimes of the molecules can span a broad range. This circumstance makes it almost impossible to analyze such cascade reactions spectroscopically in bulk but highlights simultaneously the power of SMART-EM measurements, which deconvolute stochastic processes through the direct observation of each event. Given that we can already perform single-molecule kinetic analysis using SMART-EM,[19,43] and the continuous advancement of instrumentation and information processing technology, the microscopic study of other transient intermediates will become possible in the future.

**Supporting Information**

The Supporting Information is available free of charge on the ACS Publications website.

Additional materials and methods, supporting figures and supporting video captions (PDF)

Video S1-4 (avi)

**AUTHOR INFORMATION**

**Corresponding Authors**

d.lungerich@yonsei.ac.kr

harano@chem.s.u-tokyo.ac.jp

nakamura@chem.s.u-tokyo.ac.jp

**PRESENT ADDRESS**

† Department of Applied Physics, Tokyo University of Agriculture and Technology, Nakamachi 2-24-16, Koganei, Tokyo 184-8588, Japan

# Research Center for Advanced Measurement and Characterization, National Institute for Materials Science (NIMS), 1-1 Namiki, Tsukuba 305-0044, Japan.

## ACKNOWLEDGMENT

We thank Mr. Nobuya Mamizu (SYSTEM IN FRONTIER INC.) for development of automated cross correlation software for EM images. This research is supported by JSPS KAKENHI (JP19H05459 and JP21H01758) and the Institute for Basic Science (IBS-R026-Y1). T.S. thanks MEXT (ALPS program).